\documentclass[aps,showpacs,twocolumn,longbibliography,nofootinbib,superscriptaddress]{revtex4-1}

\usepackage{graphicx} % Include figure files
\usepackage{amssymb}
\usepackage{float}
\usepackage[T1]{fontenc}
\usepackage{epstopdf}
\begin{document}
\title{Fano-control of down-conversion in a nonlinear crystal \\ embedded with  plasmonic-quantum emitter hybrid structures}

\author{Zafer Artvin}%\thanks{zartvin@metu.edu.tr}
\affiliation{Institute  of  Nuclear  Sciences, Hacettepe University, 06800 Ankara, Turkey}
\affiliation{Department of Nanotechnology and Nanomedicine, Hacettepe University, Ankara, Turkey}
\affiliation{Central Laboratory, Middle East Technical University, 06800 Ankara, Turkey}
\author{Mehmet Gunay}%\thanks{metasgin@hacettepe.edu.tr }
\affiliation{Institute  of  Nuclear  Sciences, Hacettepe University, 06800 Ankara, Turkey}
\affiliation{Current Address: Department of Nanoscience and Nanotechnology, Faculty of Arts and Science, Mehmet Akif Ersoy University, 15030 Burdur, Turkey}
\author{Alpan Bek}%\thanks{metasgin@hacettepe.edu.tr }
\affiliation{Department of Physics, Middle East Technical University, 06800 Ankara, Turkey}
\author{Mehmet Emre Tasgin}%\thanks{metasgin@hacettepe.edu.tr }
%\email{metasgin@hacettepe.edu.tr}
\affiliation{Institute  of  Nuclear  Sciences, Hacettepe University, 06800 Ankara, Turkey}

%
%
%
%
%
% \author{Mehmet Emre Ta{s}g{i}n$^{3}$, Alpan Bek$^{4}$ and Mehmet G\"{u}nay\ $^{5}$}
%\affiliation{ $^{1}$ Department of Nanotechnology and Nanomedicine, Hacettepe University\\
%$^{2}$ Central Laboratory, Middle East Technical University, 06800 Ankara, Turkey\\
%$^{3}$ Institute  of  Nuclear  Sciences, Hacettepe University, 06800 Ankara\\
%$^{4}$ Department of Physics, Middle East Technical University, 06800 Ankara, Turkey\\
%$^{5}$ Department of Nanoscience and Nanotechnology, Faculty of Arts and Science, Mehmet Akif Ersoy University, 15030 Burdur, Turkey}
%%\break

%\onecolumngrid
\begin{abstract}
Control of nonlinear response of nanostructures via path interference effects, i.e., Fano resonances, has been studied extensively. In such studies, a frequency conversion process takes place near a hot spot. Here, we study the case where the frequency conversion process takes place \textit{along the body of a nonlinear crystal}.  Metal nanoparticle-quantum emitter dimers control the down-conversion process, taking place throughout the crystal body, via introducing interfering conversion paths. Dimers behave as interaction centers. We show that a 2 orders of magnitude enhancement is possible, on top of the enhancement due to localization effects. That is, this factor multiplies the enhancement taking place due to the field localization.
\end{abstract}
%\pacs{xxx}
\maketitle

%%%%%%%%%%%%%%%%%%%%%%%%%%  body  %%%%%%%%%%%%%%%%%%%%%%%%%%
\section{Introduction}
Spontaneous down conversion~(SDC), a second-order nonlinear process, splits an incident beam into two subfrequencies labeled as signal and idler~\cite{BOYD}. Nonlinear materials,  like barium borate~\cite{IEEE14}, LiNbO3~\cite{libo3}, are employed in generating these photon pairs due to their large nonlinearities. Down-conversion process can entangle two down-converted beams/photons in various degrees of freedom: such as continuous-variable, polarization, space, time and orbital angular momentum~\cite{LightScienceApplications9}. Entangled beam/photon pairs are essential for many fundamental quantum optics experiments~\cite{PRL10} as well as a key resource in quantum communication including cryptography~\cite{PRL11}, quantum computation~\cite{PRL12} and quantum information~\cite{euler2011spontaneous}. SDC can also be used in solar cell applications by means of reducing the energy losses with converting high energy photons into two or more lower energy photons~\cite{trupke2002improving,JAP17}.

Despite such important implementations in quantum technologies~\cite{NaturePhot_2019_IntegratedQuantumTechnologies,NaturePhot_2009_QuantumTechnologies,Nanophot_2017_PlasmonicsQuantumTechnologies}, limited efficiencies of materials in parametric-down conversion process, i.e., the conversion rate, still poses a problem in technological applications~\cite{PRL10}. Some recent works~\cite{SR18, OE19} report on improvement of the conversion efficiency via changing the geometry/structure of the nonlinear crystals. Some other experimental works employ interaction of the crystal with quantum dots~(QDs), as quantum emitters~(QEs). QDs, embedded into photonic crystal structures~\cite{qd} and into nonlinear down-converting crystals, resonant with the down-converted mode~\cite{cavitysdc,bilgelt}, are shown to enhance the parametric-down conversion process. More recent works utilize the plasmonic localization effect of metal nanoparticles. For instance, recent theoretical studies~\cite{Plasmon_th1,Plasmon_th2} envision a 5 orders of magnitude enhancement in SDC efficiency when plasmonic nano-structures are embedded into nonlinear crystals. Thus, plasmonic nanoparticles, localizing (concentrating) the field into hot spots (in the crystal medium), can help nonlinearity.
Experiments on the integration of plasmonic nanostructures into nonlinear devices, such as fiber glasses~\cite{fiber2} and photonic crystals~\cite{pc2,pc3}, also report enhanced nonlinear response, e.g., on the Raman conversion process. A second harmonic generating cavity, embedded with plasmonic nanostructures, is also experimentally demonstrated to yield enhanced sum-frequency generation~\cite{
SHGcrystal_MNPs_Nanoscale2018,MNPCrystalNonlineEnhance_ACSOmega_2017,mukherjee2011one,bigot2011linear}.

The field-localizing feature of the metal nanostructures not only enhances the nonlinear processes, but also leads to enhanced light-matter interaction at the hot spots. Quantum emitters, placed at the hot spots, couple strongly to the concentrated near-field (polarization) of the plasmon excitation, which is several orders of magnitude stronger compared to their direct coupling to the incident light. Strong interaction between a QE and a metal nanoparticle (MNP) makes path interference effects visible: Fano resonances~\cite{ridolfo2010quantum}, the plasmon-analog of electromagnetically induced transparency~(EIT)~\cite{scully1999quantum1}. Similar to EIT-like behaviors in atomic clouds~\cite{scully1999quantum1}, Fano resonances can be used to control the refractive-index~\cite{panahpour2019refraction,gunay2020continuously,yuce2020ultra,karanikolas2020strong} and nonlinear conversion processes~\cite{butet2014fano,turkpence2014engineering,selen,paspalakis2014FWM}. The origin of these interference effects, e.g., enhancement and suppression of linear and/or nonlinear response, can be demonstrated with a basic analytical model, where cancellations in the denominator of a converted amplitude result in enhancement of the processes~\cite{gunay2019fano}. Besides providing such control methods on the steady-state amplitudes,  Fano resonances can also increase the lifetime of plasmon oscillations~\cite{FR1,FR2,bilgelt,ovali2020single}, leading a further accumulation of the field intensity at the hot spots, i.e., dark-hot resonances \cite{darkhot,zhang2013coherent,zhang2014coherent}.

Such plasmonic path interference effects, where the nonlinear conversion process takes place on a local region, have already been studied extensively. In these setups both the generation of the nonlinear (up or down conversion) field and the interaction of the quantum emitter with such a nano-converter take place at the hot spot. In systems, where nanoparticles are embedded into nonlinear crystals~\cite{plasnap,gianten,cuzn,libo3}, however, nonlinear process takes place all over the crystal body, i.e., not merely on the nanoparticle hot spot. 

In this paper, we study the nonlinear response of a down-converting crystal into which MNP-QE dimers are embedded, see Fig. 1a.~MNPs behave as interaction centers. They make the unlocalized down-converted field concentrate into the hot spots where a stronger interaction with a quantum emitter takes place, e.g., compared to QE-embedded crystals~\cite{plasnap,gianten,cuzn,libo3}. We show that such a setup can enhance the down-converted field 2 orders of magnitude. It should be emphasized that, this enhancement comes as a further multiplication factor on top of the MNP hot spot enhancement.

We treat the system with a basic model which has proven its compliance. The model fits almost perfectly with the exact solutions of Maxwell equations in a second harmonic~(SH) crystal cavity~\cite{gunay2019fano}, where SH conversion is implemented in the Maxwell equations. In the program we use in our Maxwell equation simulations (and in many others), however, PDC is not implemented exactly.

The paper is organized as follows. In Sec.~\ref{sec:setup}, we introduce the setup for enhancing the SDC  process. In Sec.~\ref{sec:Hamil}, we describe the physics for SDC taking place in a nonlinear crystal in which MNP-QD dimers are embedded. We define the Hamiltonian and derive the equations of motion. We obtain steady-state amplitudes for the fields. In Sec.~\ref{sec:results}, we present the parameters where 2 orders enhancement, multiplying the localization enhancement, can be achieved. We observe that, interestingly, small coupling factors can be sufficient. Sec.~\ref{sec:conclusion} contains our conclusions.

%%%%%%%%%%%%%%%%%%%%%%%%%%%%%%%%%%%%%%%%%%%%%%%%%%%%%%%%%%%%%%%%%%%%%%%%%%%%%%%%%%%%%%%%%%%%%%%%%%%%%%%%%%%%%%%%%%%%%%%%%%%%%%%%%%%%%%%%%%
%%%%%%%%%%%%%%%%%%%%%%%%%%%%%%%%%%%%%%%%%%%%%%%%%%%%%%%%%%%%%%%%%%%%%%%%%%%%%%%%%%%%%%%%%%%%%%%%%%%%%%%%%%%%%%%%%%%%%%%%%%%%%%%%%%%%%%%%%%
%%%%%%%%%%%%%%%%%%%%%%%%%%%%%%%%%%%%%%%%%%%%%%%%%%%%%%%%%%%%%%%%%%%%%%%%%%%%%%%%%%%%%%%%%%%%%%%%%%%%%%%%%%%%%%%%%%%%%%%%%%%%%%%%%%%%%%%%%%
\section{Setup} \label{sec:setup}

A laser of frequency $\omega$ pumps the $\Omega_1$ cavity mode of the nonlinear crystal, see Fig.~1. The $\omega$ oscillations in the $\Omega_1$ cavity mode, $\sim e^{-i\omega t}$, are down-converted into $\omega_2=0.3\omega$ and $\omega_3=0.7\omega$ oscillations in the cavity modes $\Omega_2$ and $\Omega_3$, respectively (see Fig.~1b.). We make the cavity field $\omega_2=0.3\omega$ interact with MNP's plasmon mode ($\Omega_p$) and introduce path interference in this converted mode. We examine the change in the amplitude of the down-converted field, $\sim e^{-i\omega_2 t}$, for different coupling parameters, i.e., between $\Omega_2$ cavity mode and the MNP, $g$, and between the MNP and the QE, $f_c$, see Fig.~2. We define the enhancement factors by comparing the $\omega_2=0.3 \omega$ down-converted intensity with and without the presence of the MNP-QE dimers.

\begin{figure}
\centering
\includegraphics[width=10cm, height=6cm]{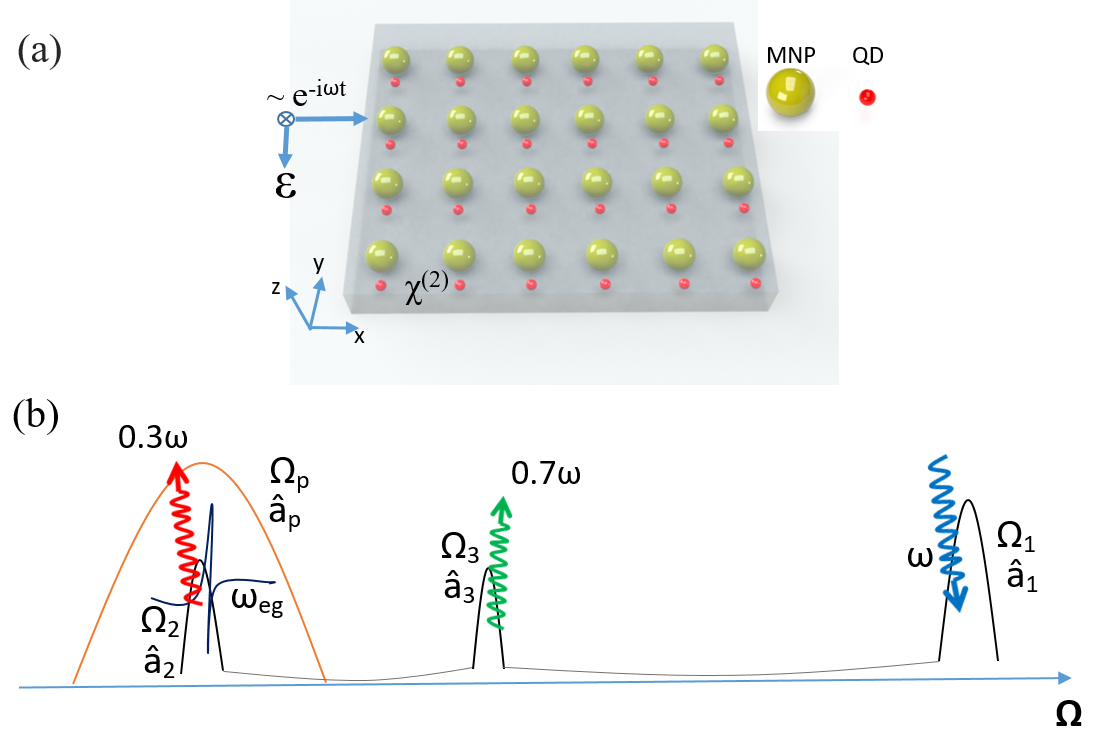}

\label{Sketch}
\caption{(a) A nonlinear crystal performs down conversion process. Metal nanoparticle~(MNP)-quantum dot~(QD) dimers are embedded into the crystal for controlling the conversion efficiency. The MNPs behave as interaction centers. They localize the already generated $\omega_2=0.3\omega$ field and introduce path interference effects. This controls the amplitude $\alpha_2$ of the $\omega_2=0.3\omega$ field. (b) Cavity modes of the nonlinear crystal $\Omega_{1,2,3}$ supports the pumped, $\omega_1=\omega$, and the down converted $\omega_2=0.3\omega$, $\omega_3=0.7\omega$ oscillations. Down-converted $\omega_2 = 0.3\omega$ photons interact with the MNP-QE hybrid structures. $\Omega_p$ is the resonance of MNP plasmons. Other, possible, modes not involved in the PDC process are not depicted. Polarization of the pump $\propto \varepsilon e^{-i\omega t}$ field, determining also the polarization of the down-converted field, is chosen along the dimer axis.}

\end{figure} 

\begin{figure}[h]
\centering
\includegraphics[width=7 cm, height=3.5 cm]{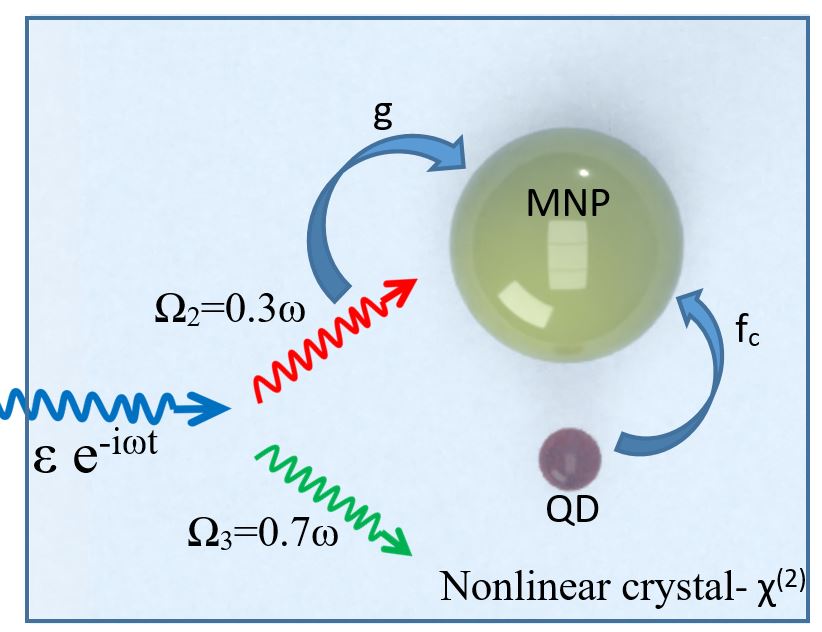}
\label{Sketch}
\centering
\caption{A sketch demonstrating the down-conversion and the interaction processes taking place in the nonlinear crystal. The input laser pumps the $\omega$ oscillations in the $\hat{a}_1$ (crystal) cavity mode. $\omega$ oscillations in the cavity is down-converted into $0.3\omega$ ($\Omega_2$ mode) and $0.7\omega$ ($\Omega_3$ mode) oscillations in the crystal. The $0.3\omega$ oscillations couple (strength $g$) with the MNP whose plasmon mode ($\Omega_p$) is around $0.3\omega$. A quantum dot~(QD), at the hot spot of the MNP, couples to the plasmon oscillations, at the frequency of $0.3\omega$. Direct coupling of the QD to the $\Omega_2$ mode, $0.3\omega$ field, is small compared to its coupling to $0.3\omega$ oscillations over the hot spot.}
\end{figure}

We introduce a basic analytical model including the damping rates of the three cavity modes, $\gamma_{1,2,3}$, and the one for the MNP plasmon mode $\gamma_p$. We use realistic values for the damping rates one can observe in typical experiments. This is similar for coupling strengths $g$ and $f_c$. The setup considered in Fig.~1a, can be manufactured by ion implantation techniques, where accelerated MNPs and QDs (or diamond  color centers) are targeted into the nonlinear crystal. A regular orientation of the MNP-QE dimers, as depicted in Fig.~1a, is not achievable using this technique~\cite{SHGcrystal_MNPs_Nanoscale2018,MNPCrystalNonlineEnhance_ACSOmega_2017,mukherjee2011one,bigot2011linear}, definitely. Hence, we also discuss an alternative and  more applicable setup in Sec.~\ref{sec:alternative}.

%%%%%%%%%%%%%%%%%%%%%%%%%%%%%%%%%%%%%%%%%%%%%%%%%%%%%%%%%%%%%%%%%%%%%%%%%%%%%%%%%%%%%%%%%%%%%%%%%%%%%%%%%%%%%%%%%%%%%%%%%%%%%%%%%%%%%%%%%%
%%%%%%%%%%%%%%%%%%%%%%%%%%%%%%%%%%%%%%%%%%%%%%%%%%%%%%%%%%%%%%%%%%%%%%%%%%%%%%%%%%%%%%%%%%%%%%%%%%%%%%%%%%%%%%%%%%%%%%%%%%%%%%%%%%%%%%%%%%
%%%%%%%%%%%%%%%%%%%%%%%%%%%%%%%%%%%%%%%%%%%%%%%%%%%%%%%%%%%%%%%%%%%%%%%%%%%%%%%%%%%%%%%%%%%%%%%%%%%%%%%%%%%%%%%%%%%%%%%%%%%%%%%%%%%%%%%%%%
\section{Hamiltonian and Equations of Motion} \label{sec:Hamil}

In this section, we first derive an effective Hamiltonian for a down-converting crystal in which MNP-QE hybrid structures are embedded. We then obtain equations of motion for the system and time evolve the equations to obtain the steady-state values of the plasmon and down-converted fields.

The setup we consider is depicted in Fig. 1. An incident light of frequency $\omega$ interacts with the nonlinear crystal and excites the $\hat{a}_1$-mode of the crystal. Down conversion process, $0.3
\omega$ and $0.7\omega$, takes place into two other cavity modes $\hat{a}_2$ and $\hat{a}_3$, respectively. The resonances of the cavity modes $\hat{a}_{1,2,3}$ are $\Omega_{1,2,3}$. The low-energy down-converted field $0.3\omega$, in the $\hat{a}_2$ mode, interacts with the embedded MNPs and excites surface plasmons ($\hat{a}_p$ mode). The interaction strength between the $\hat{a}_2$ mode and the MNP is $g$. A QE with level spacing $\omega_{eg}$ is placed at the hot spot of the MNP. The level-spacing of QE is chosen close to the plasmon field oscillations ($ \omega_{eg}  \sim \Omega_p$). Thus, QE interacts only with the $\hat{a}_2$ mode. Here, due to the enhanced hot spot intensity, there appears a strong interaction between the MNP and the QE ($f_c$). The interaction of the QE with the $\hat{a}_2$ mode, the $0.3~\omega$ down-converted field, takes place over the intense MNP hot spot, thus, the direct interaction between the $\hat{a}_2$ mode with the QE is negligible~\cite{turkpence2014engineering}, see Fig. 2. 

Hamiltonian for the down-conversion process, taking place in the nonlinear crystal medium, can be written as
 \begin{eqnarray}
 \label{e:barwig}
\hat{\cal{H}}_{\rm dc}=  \hbar\int  d^3r \chi^{(2)}(\vec{r}) \Bigl[ E_1(\textbf{r}) E_2^\ast(\textbf{r}) E_3^\ast(\textbf{r}) \: \hat{a}_3^\dagger\hat{a}_2^\dagger \hat{a}_1   
\\
+ E_1^\ast(\textbf{r}) E_2(\textbf{r}) E_3(\textbf{r}) \: \hat{a}_1^\dagger \hat{a}_2 \hat{a}_3  \Bigr ]\nonumber
 \end{eqnarray}
where $  E_1^\ast(\textbf{r})$ ($ E_2(\textbf{r}) $ $E_3(\textbf{r}) $) are the positive (negative) frequency parts of the electric fields. $ \hat{a}_j^{\dagger} $ ($ \hat{a}_j $) is the creation (annihilation) operator of the j-th mode. $ \chi_{2}= \int d^3r \: \chi^{(2)}(\vec{r}) E_1^\ast(\textbf{r}) E_2(\textbf{r})  E_3(\textbf{r})$ is an overlap integral, which determines the strength of the down conversion process~\cite{selen}. Here, one can consider $\chi^{(2)}(\bf {r})$  as a 3D step function which is zero outside the crystal body.

The total hamiltonian of the system can be written as the sum of the energies 
\begin{equation}
\hat{\cal{H}}=\hat{\cal{H}}_0 + \hat{\cal{H}}_{L} + \hat{\cal{H}}_{\rm dc} + \hat{\cal{H}}_{\rm int}.
\end{equation} 
\begin{eqnarray}
\hat{\cal{H}}_0&=& \sum_{i=1}^{3} \hbar\Omega_{i} \hat{a}_{i}^\dagger \hat{a}_i   +\hbar \Omega_p \hat{a}_{p}^\dagger \hat{a}_p
+\hbar \omega_{eg} |e \rangle \langle e|  \label{Ham0}
\end{eqnarray}
 is the total of the energies of the crystal oscillations, i.e., the pumped~($ \Omega_1 $) and the down converted~($\Omega_2$ and $\Omega_3$) crystal fields, the plasmon oscillations~($ \Omega_p $), and the two-level QE~($ \omega_{eg} $), respectively. $|g \rangle$~($|e \rangle $) is the ground (excited) state of the QE.
\begin{eqnarray}
\hat{\cal{H}}_{L}&=&i\hbar (\varepsilon  \hat{a}_1^\dagger e^{-i\omega t} -\textit{H.c.}) \label{Ham0b}
\end{eqnarray} 
is the interaction of the pump laser with the fundamental ($\hat{a}_1$) mode. Laser induces $\sim e^{-i\omega t}$ oscillations in the $\hat{a}_1$ mode. The nonlinear process 
\begin{eqnarray}
\hat{\cal{H}}_{\rm dc} &=& \hbar \chi_{2} (\hat{a}_2^\dagger \hat{a}_3^\dagger \hat{a}_1 + \hat{a}_1^\dagger \hat{a}_2 \hat{a}_3 )
\end{eqnarray}
performs the frequency conversion: converts a $\hat{a}_1$ photon into a $\hat{a}_2$ and $\hat{a}_3$ photons. Interaction of the down-converted $\omega_2=0.3\omega$ field with the MNP and the coupling of the MNP plasmon $\hat{a}_p$ (hot spot) with the QE is given by
\begin{eqnarray}
\hat{\cal{H}}_{\rm int}= \hbar g(\hat{a}_p^\dagger \hat{a}_2+\hat{a}_2 ^\dagger\hat{a}_p) + \hbar f_c (|e\rangle \langle g| \hat{a}_p  + \hat{a}_p^\dagger |g\rangle \langle e|), 
\end{eqnarray}
respectively. Coupling strength $f_c$, in units of frequencies, depends on the relative positions of the MNP and QE. $g$ depends on the MNP-size and overlap of MNP position with the $E_2({\bf r})$ cavity mode profile.

Dynamics of the system can be obtained using the Heisenberg equations of motion ($i\hbar\dot{\hat{a}}_i=[\hat{a}_i,\hat{\cal H}] $). Since we are interested in the intensities only, but do not aim to calculate the correlations, we replace the operators $\hat{a}_i$ and $ \hat{\rho}_{ij}= |i\rangle \langle j|$ with complex (c-) numbers  ${\alpha}_i$ and $ {\rho}_{ij} $ respectively~\cite{premaratne2017theory}, e.g., $\hat{a}_i\to ~\alpha_i$. The equations of motion can be obtained as
\begin{eqnarray}
\dot{{\alpha}_1}&=&-(i\Omega_1+\gamma_{1}) {\alpha}_1-i  \chi_{2} {\alpha}_2 {\alpha}_3 + \varepsilon e^{-i\omega t}\label{EOMa},\\
\dot{{\alpha}_2}&=&-(i\Omega_2+\gamma_{2}) {\alpha}_2-i  \chi_{2} {\alpha}_1 {\alpha}_3^{\ast} - i g {\alpha}_{p} \label{EOMb}, \\
\dot{{\alpha}_3}&=&-(i\Omega_3+\gamma_{3}) {\alpha}_3-i  \chi_{2} {\alpha}_1 {\alpha}_2^{\ast} \label{EOMc}, \\
\dot{{\alpha}}_p &=&  -(i \Omega_p+\gamma_{p}){\alpha}_p-i g {\alpha}_2-if_c{\rho}_{ge} \label{EOMd} ,\\
\dot{{\rho}}_{ge} &=&  -(i \omega_{eg}+\gamma_{eg}){\rho}_{ge}+i f_c{\alpha}_p({\rho}_{ee}-{{\rho}}_{gg}) \label{EOMe},\\
\dot{{{\rho}}}_{ee} &=& -\gamma_{ee} {{\rho}}_{ee}+i (f_c {\rho}_{ge} {\alpha}^\ast_p- \textit{c.c}),
\label{EOMf}
\end{eqnarray}
where $\gamma_{1,2,3}$, $ \gamma_{p} $ and $ \gamma_{ee} $  are the damping rates of  crystal modes, the plasmon mode and the quantum emitter, respectively. There exists also a constraint for the probability conservation, i.e., $ {{\rho}}_{ee} +{{\rho}}_{gg} =1$. $ \gamma_{eg} = \gamma_{ee}/2 $ is the off-diagonal decay rate for a single QE. Fields can be found by inserting the steady-state amplitudes $\alpha_1= \tilde{\alpha}_1 e^{-i\omega t} $, $\alpha_2= \tilde{\alpha}_2 e^{-i\omega_2 t} $, $\alpha_p= \tilde{\alpha}_p e^{-i\omega_2 t} $, $\rho_{eg}(t)=\tilde{\rho}_{eg} e^{-i\omega_2 t}$, $\rho_{ee,gg}=\tilde{\rho}_{ee,gg}$ and $\alpha_3= \tilde{\alpha}_3 e^{-i\omega_3 t} $  into Eqs.~(\ref{EOMa})-(\ref{EOMf}).

\begin{eqnarray}
0&=&-[i(\Omega_1-\omega)+\gamma_{1}] \tilde{\alpha}_1-i  \chi_{2} \tilde{\alpha}_2 \tilde{\alpha}_3 + \varepsilon \label{steayEqa},\\
0&=&-[i(\Omega_2-\omega_2)+\gamma_{2}] \tilde{\alpha}_2-i  \chi_{2} \tilde{\alpha}_1 \tilde{\alpha}_3^{\ast} - i g \tilde{\alpha}_{p} \label{steayEqb}, \\
0&=&-[i(\Omega_3-\omega_3)+\gamma_{3}] \tilde{\alpha}_3-i  \chi_{2} \tilde{\alpha}_1 \tilde{\alpha}_2^{\ast} \label{steayEqc}, \\
0&=&  -[i( \Omega_p-\omega_2)+\gamma_{p}]\tilde{\alpha}_p-i g \tilde{\alpha}_2-if_c\tilde{\rho}_{ge} \label{steayEqd} ,\\
0&=&  -[i( \omega_{eg}-\omega_2)+\gamma_{eg}]\tilde{\rho}_{ge}+i f_c\tilde{\alpha}_p(\tilde{\rho}_{ee}-{\tilde{\rho}}_{gg}) \label{steayEqe},\\
0&=& -\gamma_{ee} {\tilde{\rho}}_{ee}+i (f_c \tilde{\rho}_{ge} \tilde{\alpha}^\ast_p- \textit{c.c}),
\label{steayEqf}
\end{eqnarray}
reminding that $\omega_2=0.3\omega$ and $\omega_3=0.7\omega$.

For an upconverting process, e.g., second harmonic generation~(SHG) we are able to obtain a simple intuitive expression for the converted field amplitude~\cite{gunay2019fano} (see the expression in Appendix for a second harmonic converting cavity). There, we can use that expression to anticipate the presence of enhancement and suppression phenomena which matches very well with the 3D solutions. Unfortunately, here we could not manage to obtain such simple expression for the down conversion process, so that we cannot demonstrate the reader explicitly how path interference, i.e., cancellations, emerge in the SDC process.

%%%%%%%%%%%%%%%%%%%%%%%%%%%%%%%%%%%%%%%%%%%%%%%%%%%%%%%%%%%%%%%%%%%%%%%%%%%%%%%%%%%%%%%%%%%%%%%%%%%%%%%%%%%%%%%%
%%%%%%%%%%%%%%%%%%%%%%%%%%%%%%%%%%%%%%%%%%%%%%%%%%%%%%%%%%%%%%%%%%%%%%%%%%%%%%%%%%%%%%%%%%%%%%%%%%%%%%%%%%%%%%%%
%%%%%%%%%%%%%%%%%%%%%%%%%%%%%%%%%%%%%%%%%%%%%%%%%%%%%%%%%%%%%%%%%%%%%%%%%%%%%%%%%%%%%%%%%%%%%%%%%%%%%%%%%%%%%%%%
\section{Enhancement of Down-Conversion} \label{sec:results}
 The enhancement factor~(EF) for the down-converted field intensity,
\begin{eqnarray}
{\quad EF}=\frac{|\alpha_2 (f_c\neq 0,g\neq 0)|^2}{|\alpha_2(f_c=0,g = 0)|^2} 
\label{EF}
\end{eqnarray}
is defined as the ratio of the down converted intensities in the presence~($ f_c\neq 0,g\neq 0 $) and absence ($ f_c=0,g=0 $)  of the MNP-QE hybrid structure. We compare the case (i) when MNP-QE dimer is present versus (ii) no particle is present (a bare crystal). We calculate the enhancement factors at the steady-state, from the time evolution Eqs. (\ref{EOMa})-(\ref{EOMf}) and using Eq.~(\ref{EF}). We investigate the effect of MNP-QE hybrid structure on the output signal of the down converting crystal. %Moreover, plasmon field at MNP ($\alpha_p$) for  both MNP-QE hybrid system or only presence of MNP in nonlinear crystal can be calculated.

\begin{figure}[h]
\begin{center} 
%\begin{subfigure}[b]{1\linewidth}
 \includegraphics[width=9cm, height=4.5cm]{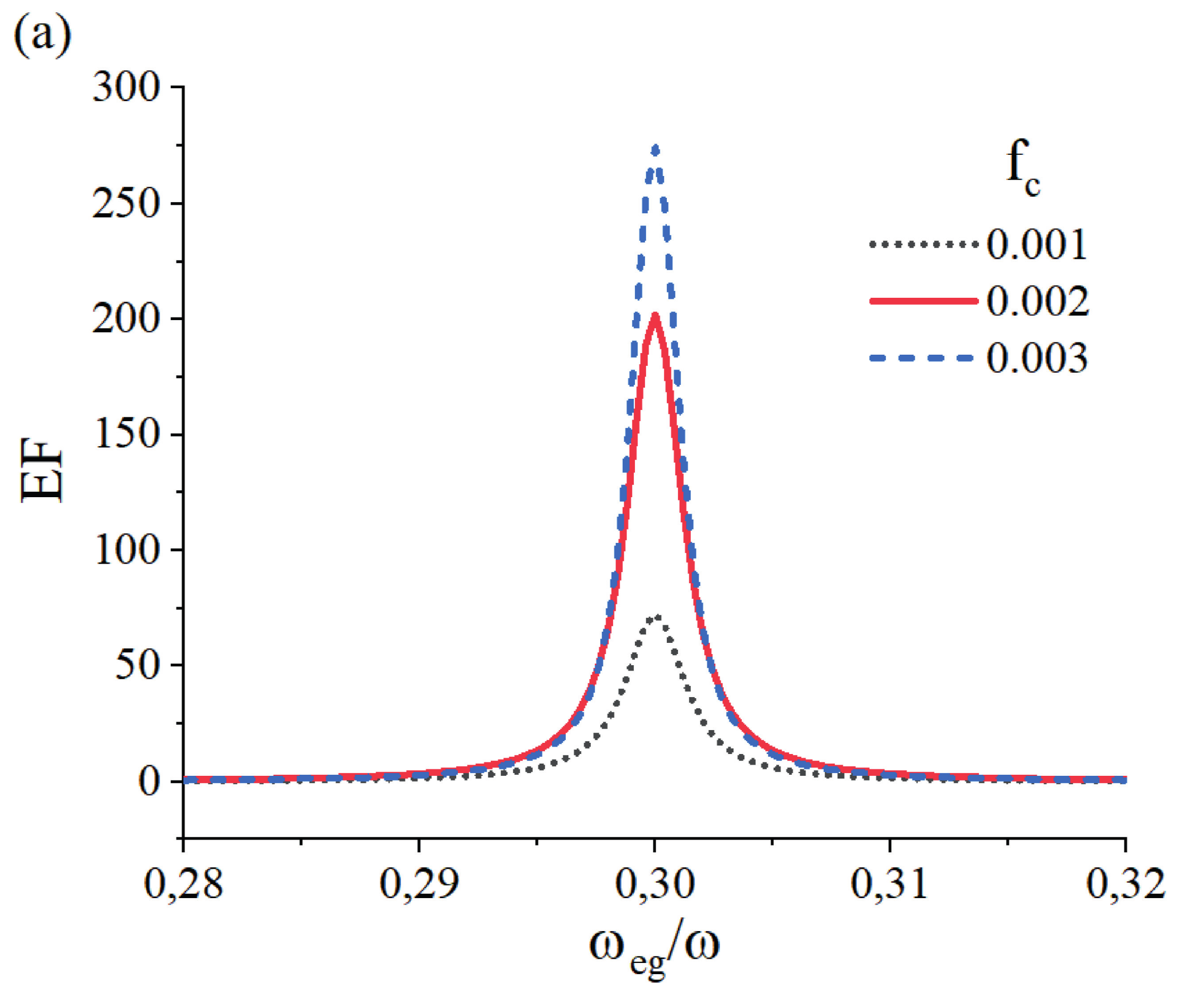}
%\caption{}

%\end{subfigure}
%\begin{subfigure}[b]{1\linewidth}
\includegraphics[width=8.4cm, height=4.5cm]{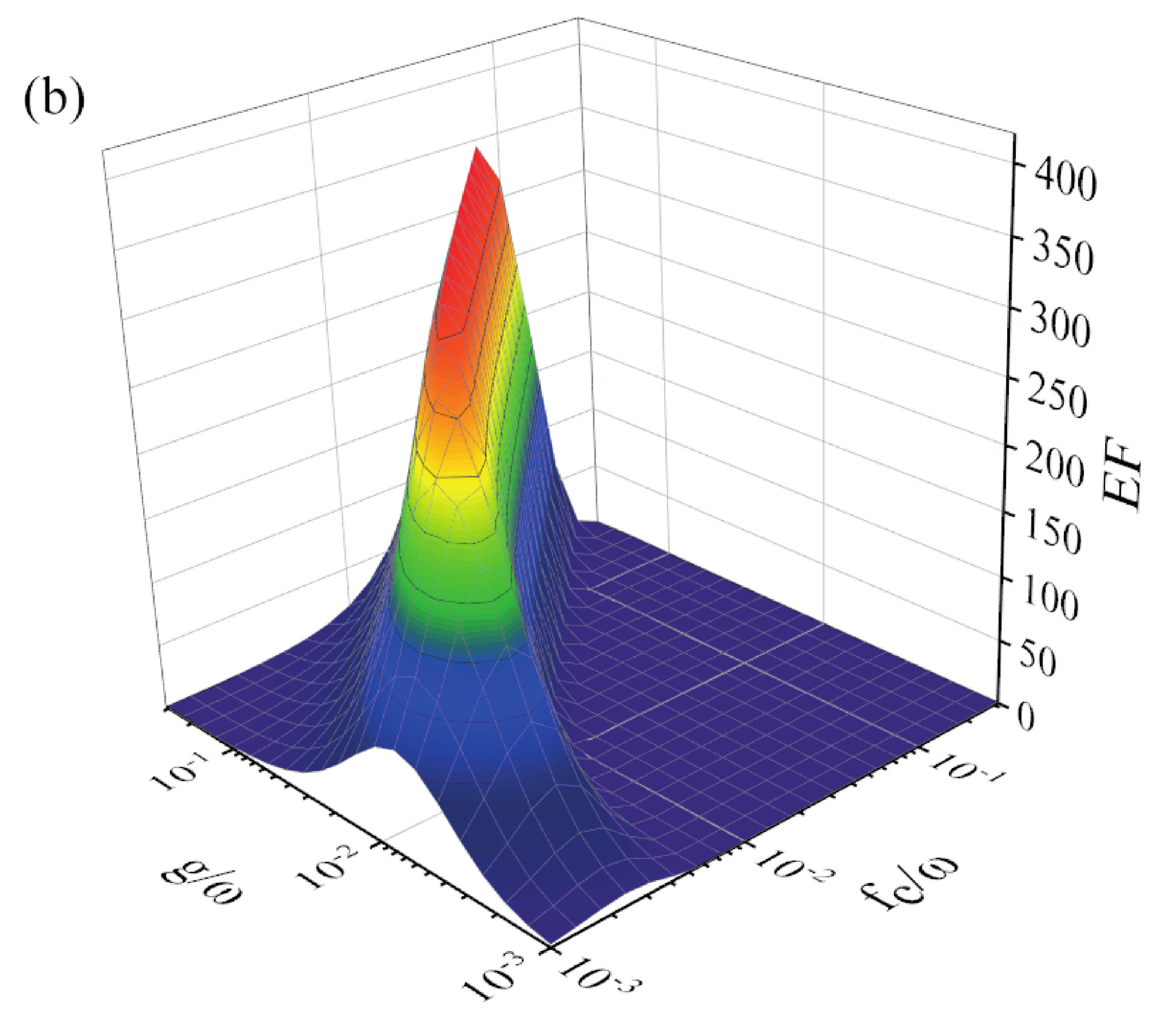}
%\caption{}
%\end{subfigure}
%\begin{subfigure}[b]{1\linewidth}
%\includegraphics[width=0.55\linewidth]{fig3c.png}
%\caption{}
%\end{subfigure}
\caption{Enhancement factor~(EF) of the down converted $0.3\omega$ signal intensity (a) versus QE's level-spacing $\omega_{eg}$ calculated for different MNP-QE interaction strengths  $f_c$. We set $g=0.01\omega$ for the coupling of the MNP to the $\alpha_2$ crystal field. $\alpha_2$ supports the $0.3\omega$ oscillations. (b) EF, at fixed $\omega_{eg}=0.3\omega$, calculated for different $f_c$ and $g$ couplings.  There appears 300 to 400 fold EFs, crudely, for the ratios $f_c/g\sim 1/4$. Other parameters we use are $\Omega_1=\omega$, $\Omega_2=0.3\omega$, $\Omega_3=0.7\omega$, $\Omega_p= 0.3\omega$, $\gamma_1=\gamma_2=\gamma_3=5\times10^{-4}\omega$, $\gamma_{p}=0.1~\omega$, $\gamma_{eg}=10^{-5}~\omega$ and $  \chi_{2}=2\times 10^{-9}~\omega$.}
\end{center}
\end{figure}

We examine the down converted $\omega_2=0.3\omega$ field intensity, $|\alpha_2|^2$, for different interaction strengths $g$ and $f_c$. These parameters change with MNP size and the position of the QE. Additionally, in the alternative setup given in Fig.~\ref{fig2} in Sec.\ref{sec:alternative}, QD level spacing $\omega_{eg}$ can also be continuously tuned via an applied voltage. In Fig.~\ref{fig2}, MNPs and QDs are positioned on the surface of the crystal and they are coupled to the down converted field over the evanescent waves.

Fig. 3a shows that path interference phenomenon can achieve 2-orders of magnitude enhancement factors in the presence of the MNP-QD dimers, despite the fact that the damping rate of the MNP $\gamma_p$ (coupled to the $\hat{a}_2$ mode) is orders of magnitude larger than the one for the $\alpha_2$ crystal mode. We scale the frequencies with $\omega$, the frequency of the laser pumping the $\alpha_1$ crystal mode. We use $g=0.01\omega$, $\Omega_1=\omega$, $\Omega_2=0.3\omega$, $\Omega_3=0.7\omega$, $\Omega_p= 0.3\omega$, $\gamma_1=\gamma_2=\gamma_3=5\times10^{-4}\omega$, $\gamma_{p}=0.1\omega$  and $\gamma_{eg}=10^{-5}\omega$\cite{gunay2019fano}. We consider a small (arbitrary) second order nonlinear (overlap) factor  $\chi_{2}=2\times 10^{-9}\omega$ whose actual value does not change the enhancement ratios.

Fig. 3b demonstrates how the strength of the interaction (i) between the down-converted $\alpha_2$ field and the MNP~($g$) and (ii) between the MNP plasmon (hot spot) field and the QE~($ f_c $)  affect the enhancement factor for a fixed QE level spacing of ($ \omega_{eg}=0.3\omega$). We observe that a substantial enhancement takes place at a certain ratio of $f_c/g$, subject to the condition $f_c<g$. We can crudely observe this ratio as $f_c/g\sim 1/4$, where enhancement of the $|\alpha_2|^2$ down-converted intensity appears~\footnote{No approximations are carried out in the solutions of the equations of motion. The presented results are the time evolution, i.e., exact solutions, of equations~(\ref{EOMa}-\ref{EOMf}).}. As the $f_c$ value gets closer to the $g$, for $f_c>g$, enhancement factor (EF) decreases.
\begin{figure}[h]
 \centering
\includegraphics[width=8.4cm, height=4.5cm]{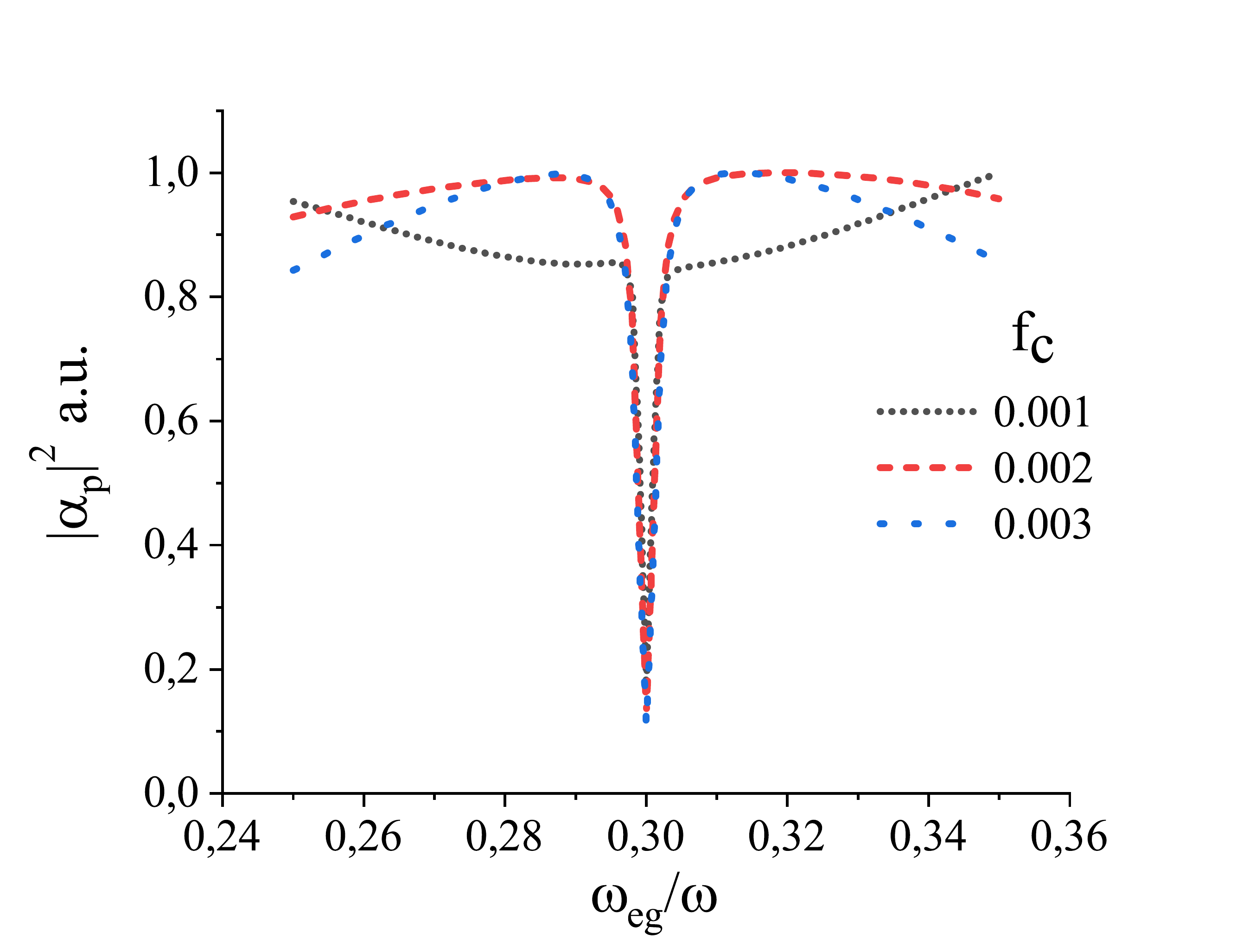}
\caption{MNP plasmon intensity ($|\alpha_p|^2$) versus quantum emitter level-spacing $\omega_{eg}$ calculated for different values of the MNP-QE coupling strength $f_c$. We observe the suppression of the plasmon field at $\omega_{eg}=0.3\omega$. Other parameters are $g=0.01\omega$, $\gamma_p=0.1\omega$,  $\Omega_1=\omega$, $\Omega_2=0.3\omega$, $\Omega_3=0.7\omega$, $\Omega_p= 0.3\omega$, $\gamma_1=\gamma_2=\gamma_3=5\times 10^{-4}~\omega$, $\gamma_{eg}=10^{-5}\omega$,  and $\chi_{2}=2 \times 10^{-9}~\omega$.  } 
\end{figure} 

In Fig. 4, we plot the MNP plasmon field intensity ($|\alpha_p|^2$) versus level-spacing of the QE, $\omega$, for different $f_c$ values. The plasmon intensity demonstrates a dip at $\omega_{eg}$= 0.3$\omega $.

%We calculated the down converted field amplitude with respect to the various resonance frequencies of MNP for the system that nonlinear crystal contains only MNPs without QEs.
%The $ \alpha_2$ signal  level of the system decrease at the plasmon resonance frequency of $ \omega_p = 0.3~\omega $ as depicted in Fig. 3c. These results show that the path interference effects produce $\alpha_2 $ signal enhancements and not originated from the localization effects. As mentioned above our model does not cover enhancement due to  localization effects.

%
%%nonlinear phenomena second-harmonic generation, the nonlinear field could be expressed analytically. According to these formulas, we could say that the origin of such enhancement and suppression come from the path interference effects at the non-linear response since the presence of MNP-QE defines extra terms which may lead to constructive or destructive interferences. When $\omega_{eg}$ of the QE or the interaction strengths are tuned properly ($\omega_{eg}$ = 2$\omega$) enhancement of down-converted field could bee seen besides suppression at plasmon field. Depending on the parameters, better cancellation can be made, and larger enhancement at the output signal can be obtained~\cite{selen}.

%%%%%%%%%%%%%%%%%%%%%%%%%%%%%%%%%%%%%%%%%%%%%%%%%%%%%%%%%%%%%%%%%%%%%%%%%%%%%%%%%%%%%%%%%%%%%%%%%%%%%%%%%%%%%%%%%%%%%%
\subsection*{ Enhancement near $\omega_{eg}\simeq \omega_2$}

When Figs. 3a and 4 are compared, one can realize that a 300 fold enhancement appears in  the down-converted field $|\alpha_2|^2$, for $\omega_{eg} = 0.3\omega$, when the plasmon mode of the MNP is suppressed. Thus, it is natural to get suspicious if this 300 fold enhancement occurs, actually, due to the suppression of the plasmon excitation around $\omega_{eg}= 0.3\omega$ , i.e., when strong absorption of the MNP is turned off.  In other words, in the calculation of the EF~(\ref{EF}), are we ``mistakenly'' comparing $|\alpha_2(g\neq 0, f_c\neq 0)|^2$ with $|\alpha_2(g\neq 0)|^2$? We check that we are calculating the correct EF, i.e., defined in Eq.~(\ref{EF}). 

%We, such a suspicion in mind, checked our calculations several times. We inspected if we define the enhancement factor of $|\alpha_2|^2$, as a mistake, by comparing the two cases, i.e., when (i) MNP-QE dimer is present versus (ii) MNP is present alone. We confirmed that in our all results we compare the case (i) when MNP-QE dimer is present versus (ii) no particle is present (a bare crystal), i.e., the one in Eq.~(11). %

In Fig 5, we test the suppression phenomenon by using different parameters. After carrying out a neat inspection for the new parameters, we realize that, the maximum of the enhancement does not appear at $\omega_{eg}=0.3\omega$, but it appears at $\simeq 0.297\omega$ in Fig.~5a. In Fig.~5b, however, we observe that plasmon excitation is suppressed maximally still at $\omega_{eg}=0.3\omega$. For other choices of the parameters $f_c$ and $g$, this discrepancy is more explicitly. That is, maximum $|\alpha_2|^2$ enhancement appears for an $\omega_{eg}$ which is apparently different than the $\omega_{eg}=0.3\omega$ where plasmon absorption is suppressed.
\begin{figure}[h]
 \centering
\includegraphics[width=8.4cm, height=4.5cm]{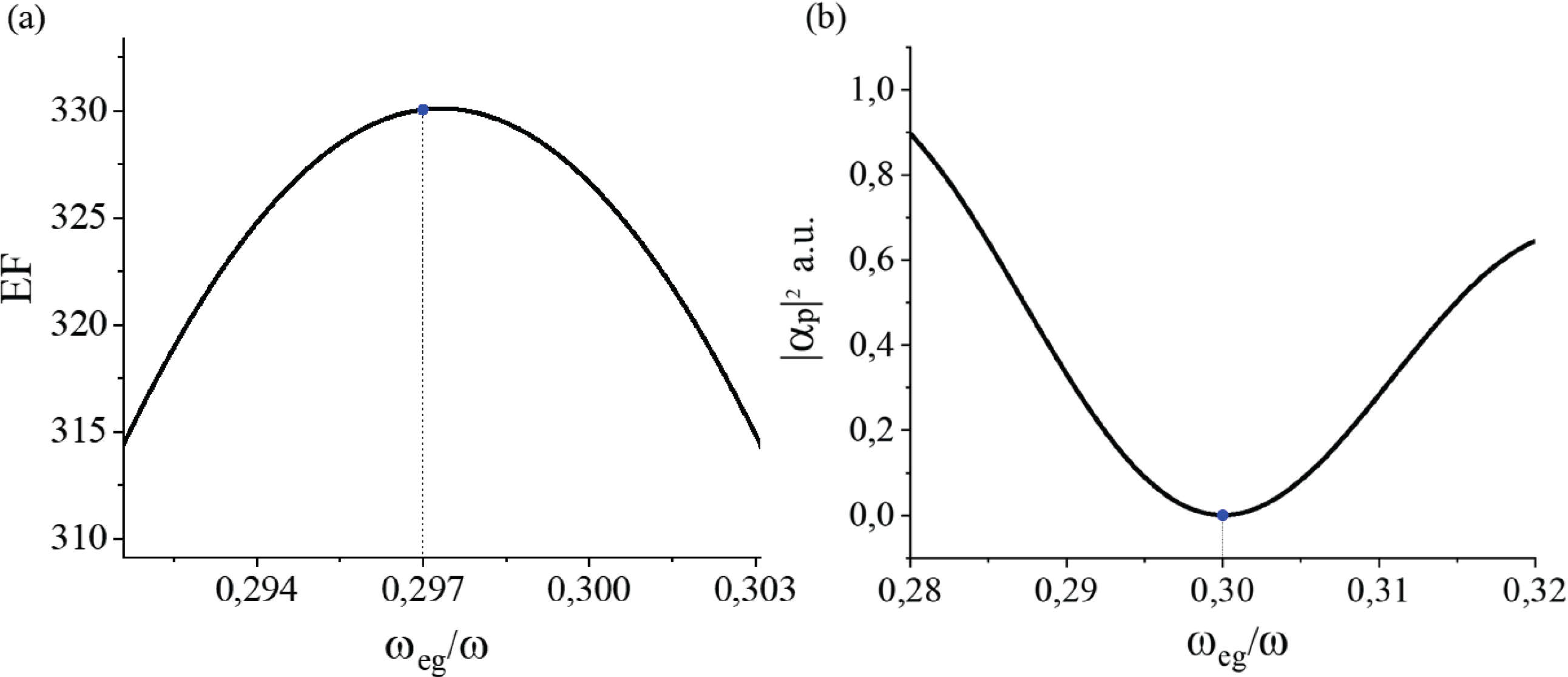}
\caption{ (a) Maximum enhancement factor is observed at $\omega_{eg}\simeq 0.297\omega$ using different parameters. (b) Whereas, the maximum suppression of the MNP excitation is still at $\omega_{eg}=0.3\omega$ for the new parameter set $\gamma_p=0.01~\omega$, $f_c = 0.03\omega$, $g=0.1\omega$, $\Omega_1=\omega$, $\Omega_2=0.3\omega $, $\Omega_3= 0.7\omega$, $\Omega_p= 0.3\omega$, $\gamma_1=\gamma_2=\gamma_3=5\times 10^{-4}\omega$, $\gamma_{eg}=10^{-5}\omega$,  and $\chi_{2}= $  $2\times 10^{-9}\omega$. } 
\end{figure}

Actually, in Ref.~\cite{gunay2019fano} we are able to obtain a simple expression for the steady-state amplitude of SHG and on this expression we are able to explain why conversion enhancement takes place near $\omega_{eg} \simeq \omega_{\rm \scriptscriptstyle NL}=2\omega$. In Ref.~\cite{gunay2019fano}, see the Appendix, we realize that the term $(\ldots) / [ i(\omega_{eg}-\omega_{\rm \scriptscriptstyle NL}) + \gamma_{eg} ]$ in the denominator can become significant to perform cancellations near $\omega_{eg} \simeq \omega_{\rm \scriptscriptstyle NL}=2\omega$. Please see Appendix for further explanation.

%%%%%%%%%%%%%%%%%%%%%%%%%%%%%%%%%%%%%%%%%%%%%%%%%%%%%%%%%%%%%%%%%%%%%%%%%%%%%%%%%%%%%%%%%%%%%%%%%%%%%%%%%%%%%%%%%%%%%%
\subsection*{ The idler intensity $|\alpha_3|^2$ }

The ``$\textit{steady-state}$'' intensity of the other (idler) mode,~$\alpha_3$, gets almost unaffected from the enhancement of the $\alpha_2$ mode. At first sight, this situation appears a bit counter-intuitive considering the argument (Arg.1) below. (Arg.1) Yes, the number conservation constraint, originating from the process $\hat{a}_3^\dagger\hat{a}_2 ^\dagger\ \hat{a}_1$, between the $|\alpha_2|^2$ and $|\alpha_3|^2$, is broken by the presence of the plasmonic (dimer) term in Eq. (8) and Eqs. (11-12). MNP-QE dimer couples only to the $\alpha_2$ mode. In the frequency conversion process, however, the dimer can only absorb photons from the $\alpha_2$ mode. Because the dimer is not pumped elsewhere. Thus, enhancement of $|\alpha_2|^2$ without altering the $|\alpha_3|^2$, at first may appear weird? (Arg.2) Nevertheless, after noting that $|\alpha_{2,3}|^2$ are the ``steady-state'' values for the signal and idler, i.e., not the total number of generated photons, situation turns out to be reasonable~\footnote{ As an illuminating example: at linear Fano resonances the steady-state amplitude of the plasmon excitation is suppressed \cite{limonov2017fano}. When the time evolution is considered, in contrary, Fano resonance creates much intense hot spots \cite{darkhot},\cite{chu2010double}}. %\footnotemark[1].

While here we consider QEs for longer lifetime particles, coupled to the MNPs, they can also be replaced with metal nanostructures supporting long-life dark-modes. Actually, metal nanostructures, or even a single MNP structure, can support both bright and dark modes where the two modes interact with each other and result in Fano resonances~\cite{Remo2014dark,bilgelt}.

%%%%%%%%%%%%%%%%%%%%%%%%%%%%%%%%%%%%%%%%%%%%%%%%%%%%%%%%%%%%%%%%%%%%%%%%%%%%%%%%%%%%%%%%%%%%%%%%%%%%%%%%%%%%%%%%%%%%%%%%%%%%%%%%%%%%%%%%%%
\subsection*{Retardation Effects}

We note that the analytical model, we base our results on, does not take the retardation effects and increased density of modes into account. 3D FDTD solutions of Maxwell equations show that retardation effects do not destroy the appearance of Fano resonances when the nonlinearity takes place on (or close to) a nanoparticle~\cite{selen,gunay2020controlling}. The case we study here, however, considers a nonlinearity conversion which takes place all over the nonlinear crystal, i.e., not only on the hot spot. Thus, the retardation effects have to be revisited~\footnote{\label{fn:3Dsimulations} Second harmonic process is encoded in many FDTD programs through the exact form of the nonlinear 3D Maxwell equations. The down-conversion process (neither, e.g., Raman conversion process), however, is not treated with exact solutions of nonlinear Maxwell equations, but their enhancement is predicted from the localization factors~\cite{plasnap,cuzn,gianten}.}.

Actually, in a new study~\cite{gunay2019fano}, we just recently managed to show that retardation effects do not wash out the path interference phenomenon. In Ref.~\cite{gunay2019fano}, we consider a setup, similar to the one in Fig.~1, but inspect the second harmonic generation process~\footnotemark[3]. More explicitly, solutions of 3D nonlinear Maxwell equations show that an extra enhancement due to path interference multiplies the enhancement due to localization (of MNPs).

We also note that our basic model does not account the change (increase) in the density of modes appearing for a QE near MNP. Our aim in this paper, already, is to study solely the path interference effects on the parametric down conversion process. In the present work, we concentrate on the enhancement factors taking place due to the path interferences~\footnote{3D simulations in Ref.~\cite{gunay2019fano} shows that such effects are not washed out by retardation effects.}. Our basic model cannot take into account the enhancements due to localization; thus, presents the ones due to path interference only.

%%%%%%%%%%%%%%%%%%%%%%%%%%%%%%%%%%%%%%%%%%%%%%%%%%%%%%%%%%%%%%%%%%%%%%%%%%%%%%%%%%%%%%%%%%%%%%%%%%%%%%%%%%%%%%%%%%%%%%%%%%%%%%%%%%%%%%%%%%
%%%%%%%%%%%%%%%%%%%%%%%%%%%%%%%%%%%%%%%%%%%%%%%%%%%%%%%%%%%%%%%%%%%%%%%%%%%%%%%%%%%%%%%%%%%%%%%%%%%%%%%%%%%%%%%%%%%%%%%%%%%%%%%%%%%%%%%%%%
%%%%%%%%%%%%%%%%%%%%%%%%%%%%%%%%%%%%%%%%%%%%%%%%%%%%%%%%%%%%%%%%%%%%%%%%%%%%%%%%%%%%%%%%%%%%%%%%%%%%%%%%%%%%%%%%%%%%%%%%%%%%%%%%%%%%%%%%%%
\section{An Alternative Setup} \label{sec:alternative}

\begin{figure}[h]
 \centering
\includegraphics[width=8cm, height=3.5cm]{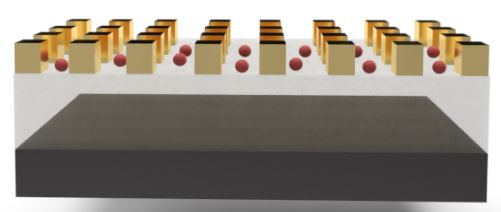}
\caption{MNP-QE dimers can be coupled to the down-converted $\omega_2 = 0.3\omega$ field via evanescent waves. Dimers can be controllably positioned on the mode lobe maxima. Controlled positioning of dimers~\cite{bek2008fluorescence} can also work in the favor of cooperative operation of path interferences from many dimers. Such a setup also enables the tuning of the quantum dot level spacing ($\omega_{eg}$) via an applied voltage.} 
\label{fig2}
\end{figure} 
An alternative and probably more feasible experimental scheme can be seen in Fig. 6. Interaction of the converted field $\omega_2=0.3\omega$ with MNP-QE dimers can also be achieved via evanescent fields. Lithography techniques~\cite{altissimo2010beam} can be used to fabricate periodic metal nanostructures on the nonlinear crystal surface and QDs can be deposited between these metal nanostructures. When the dimers are positioned on the mode lobe maxima of the converted cavity mode, evanescent radiation couples to the dimers, see Fig. 6. In such a case, the level-spacing of the quantum dots can also be tuned continuously via an applied voltage. Size, shape, type of material and the quantity of embedded nanoparticles affect the nonlinear properties of the crystal. 

In such a configuration, interaction of the down-converted $\omega_2=0.3\omega$ mode with the MNP will be smaller compared to the one in Fig.~1. Nevertheless, Fig.~3 demonstrated that relatively smaller coupling, both for $g$ and $f_c$, are sufficient compared to the path interference schemes for local frequency conversion~\cite{turkpence2014engineering,selen}.

\section{Summary and Discussions} \label{sec:conclusion}

In this work, we study path interference effects in a nonlinear crystal generating down-converted signals. A long-life quantum emitter, weakly interacting with the down-converted field alone, is made couple with the down-converted field over an MNP’s hot spot. In other words, MNPs are utilized as interaction centers. Analytical results show that it is possible to strengthen the down-converted signal more than 2 orders of magnitude in a nonlinear crystal having an inherently weak output, when the parameters are chosen accordingly. We stress that this enhancement factor further multiplies the enhancement occurring due to the localization field (hot spot) enhancement.

In recent experiments~\cite{plasnap,cuzn,gianten} and the theoretical predictions~\cite {Plasmon_th1,Plasmon_th2}, the pumped cavity mode, $\Omega_1$ mode, oscillating with $ e^{-i\omega t}$, is coupled with the MNPs. Field localization effect of MNP is utilized~\cite{plasnap,cuzn,gianten}. In the present study, in difference, the MNPs couple with the down-converted signal $\omega_2 =0.3\omega$. The present work can neither treat the enhancement due to localization. Because it does not conduct a simulation of 3D Maxwell equations. The localization effect is referred only as strong light-MNP and MNP-QE couplings.

Retardation effects, in the studied system, can be more demolishing compared to the ones for a localized nonlinear conversion process. Yet, a very recently~\cite{gunay2019fano} study demonstrates that retardation effects do not wash out the path interference phenomenon, in a second harmonic converting crystal (see also appendix). Our work presents the first demonstration of the path interference effect in a spontaneous down-converting nonlinear crystal.

While the actual aim of this work is to see the path interference effects, i.e., disassociating the effects of the localization, we can propose the following schemes for achieving much stronger spontaneous down conversion enhancements. Using a metal nanostructure, supporting two plasmon modes resonant to $\omega_1 =~\omega$ and $\omega_2=0.3\omega$, could enhance the down conversion process (i) by localizing both fields and (ii) by controlling the enhancement via path interference the QE ($\omega_{eg}$) creates around the lower-energy plasmon mode\footnote{This is a double-resonance scheme \cite{chu2010double}. A triple resonance scheme can be further considered for much stronger enhancement.}, near $\omega_2= 0.3\omega$ . Still, the path interference enhancement multiplies the localization enhancement.

Besides embedding MNP-QE dimers into the crystal, Fig. 1, they can also be coupled to the down-converted cavity mode via evanescent waves, e.g., by positioning the MNP-QE dimers to the maxima lobes, see Fig.~\ref{fig2}. This configuration also allows continuous tuning of the QE level spacing, $\omega_{eg}$, via an applied voltage for observing the path interference spectrum. The presented interference scheme, controlling ``unlocalized" nonlinear processes, necessitates much smaller coupling strengths compared to the one for local nonlinearities~\cite{turkpence2014engineering,selen}. This makes the setup in Fig.~6 feasible. 

Finally, we stress that gaining control over these phenomena or even developing an understanding on them have crucial importance in the development of new quantum technologies. 

\section*{Acknowledgements}
MET and MG acknowledges support from TUBITAK Grant No. 117F118. MET and ZA  acknowledge support from T\"{U}BA-GEB\.{I}P 2017 award.

 \renewcommand{\theequation}{A.\arabic{equation}}
  % redefine the command that creates the equation no.
  \setcounter{equation}{0}  % reset counter 
\appendix

\section*{Appendix}

In Ref.~\cite{gunay2019fano}, we study the control (in particular enhancement) of second harmonic process in a nonlinear cavity, via MNP-QE dimers, both analytically and by exact solution of 3D Maxwell equations. In that upconverting system, we are able to obtain a simple expression for the SH field amplitude, say $\beta_2$, in terms of the first harmonic amplitude
\begin{equation}
\tilde{\beta}_2= \frac{-i\chi_2}{[i(\Omega_2-2\omega)+\gamma_2]+\frac{|g|^2}{[i(\Omega_p-2\omega)+\gamma_p]- \frac{|f_p|^2y}{[i(\omega_{eg}-2\omega)+\gamma_{eg}]}}  } \tilde{\beta}_1^2. \label{appendix:1}
\end{equation}

Eq. (\ref{appendix:1}) reveals , in Ref. \cite{gunay2019fano}, how upconversion enhancement or suppression take place. Simply, (i) cancellations in the denominator (call as path interference) reduce the denominator resulting an enhanced SHG. In contrary, (ii) cancellations can also enlarge denominator at certain parameters  where a suppression in the SHG is obtained.

In Ref.~\cite{gunay2019fano}, Eq.~(\ref{appendix:1}) can envisage the exact solution of 3D Maxwell equations (FEM simulations) perfectly. Thus, one can also clearly witness that the predicted resonances are not washed out by retardation effects in the path interference control of an unlocalized process.

FEM simulations \cite{gunay2019fano} also clearly demonstrates that SHG enhancement can take place without any local field enhancement.

{\it Enhancement near $\omega_{eg}\simeq \omega_{\scriptscriptstyle \rm NL}=2\omega$.}--- In Eq.~(\ref{appendix:1}), one can see that $|f_p|^2 y/[\ldots]$ term becomes sufficiently large near $\omega_{eg}\simeq \omega_{\scriptscriptstyle \rm NL}=2\omega$. Hence, in this regime it can create a cancellation with the $[i(\Omega_p-2\omega)+\gamma_p]$ term, making $|g|^2/(\ldots)$ term sufficiently large for performing cancellation in the $[i(\Omega_2-2\omega)+\gamma_2]$ term. Consequently, the denominator of $\beta_2$ can become smaller. 

This argument explains why we obtain SHG enhancement near $\omega_{eg}\simeq \omega_{\scriptscriptstyle \rm NL}=2\omega$ in Ref.~\cite{gunay2019fano}, a phenomenon also appearing in this work. In the presented work, however, we cannot manage to obtain such an analytical expression.

\bibliography{bibliography}

%%%%%%%%%%%%%%%%%%%%%%% References %%%%%%%%%%%%%%%%%%%%%%%%%

%%%%%%%%%% If preparing manually:
% \begin{thebibliography}{1}
% \newcommand{\enquote}[1]{``#1''}

% \bibitem{Zhang:14}
% Y.~Zhang, S.~Qiao, L.~Sun, Q.~W. Shi, W.~Huang, L.~Li, and Z.~Yang,
%   \enquote{Photoinduced active terahertz metamaterials with nanostructured
%   vanadium dioxide film deposited by sol-gel method,}
%   {\protect\JournalTitle{Optics Express}} \textbf{22}, 11070--11078 (2014).

% \bibitem{OSA}
% {Optical Society}, \enquote{{OSA Publishing},}
%   \url{http://www.osapublishing.org}.

% \bibitem{FORSTER2007}
% P.~Forster, V.~Ramaswamy, P.~Artaxo, T.~Bernsten, R.~Betts, D.~Fahey,
%   J.~Haywood, J.~Lean, D.~Lowe, G.~Myhre, J.~Nganga, R.~Prinn, G.~Raga,
%   M.~Schulz, and R.~V. Dorland, \enquote{Changes in atmospheric consituents and
%   in radiative forcing,} in \enquote{Climate Change 2007: The Physical Science
%   Basis. Contribution of Working Group 1 to the Fourth assesment report of
%   Intergovernmental Panel on Climate Change,}  S.~Solomon, D.~Qin, M.~Manning,
%   Z.~Chen, M.~Marquis, K.~B. Averyt, M.~Tignor, and H.~L. Miler, eds.
%   (Cambridge University Press, 2007).

% \end{thebibliography}

\end{document}